\documentstyle[twocolumn,aps,pre,eqsecnum,epsfig]{revtex}

\newcommand{\hr}{{\bf \hat r}}
\newcommand{\br}{{\bf \bar r}}
\newcommand{\hdr}{\hat{\delta{\bf r}}}
\newcommand{\bdr}{\bar{\delta{\bf r}}}

\begin{document}
\draft 
\title{
The multifractal structure of chaotically advected chemical 
fields} 

\author{Zolt\'{a}n Neufeld$^{1,2}$, Crist\'{o}bal L\'{o}pez$^{1}$, 
Emilio Hern\'{a}ndez-Garc\'{\i}a$^1$, and Tam\'{a}s T\'{e}l$^3$}
\address{$^1$Instituto Mediterr\'{a}neo de Estudios Avanzados IMEDEA\\
CSIC-Universitat de les Illes Balears, E-07071 Palma de Mallorca, 
Spain \\ $^2$Department of Applied Mathematics and Theoretical 
Physics, University of Cambridge\\ Silver Street, Cambridge CB3 
9EW, UK\\ $^3$E\"otv\"os University, Department of Theoretical 
Physics, H-1518 Budapest, Hungary} 

\date{July 15, 1999} 
\maketitle 

\begin{abstract}
The structure of the concentration field of a decaying substance 
produced by chemical sources and advected by a smooth 
incompressible two-dimensional flow is investigated. We focus our 
attention on the non-uniformities of the H\"older exponent of the 
resulting filamental chemical field. They appear most evidently in 
the case of open flows where irregularities of the field exhibit 
strong spatial intermittency as they are restricted to a fractal 
manifold. Non-uniformities of the H\"older exponent of the 
chemical field in closed flows appears as a consequence of the 
non-uniform stretching of the fluid elements. We study how this 
affects the scaling exponents of the structure functions, 
displaying anomalous scaling, and relate the scaling exponents to 
the distribution of finite-time Lyapunov exponents of the 
advection dynamics. Theoretical predictions are compared with 
numerical experiments. 
\end{abstract}


\section{Introduction}
\label{sec:intro}

Mixing in fluids plays an important role in nature and technology 
with implications in areas ranging from geophysics to chemical 
engineering\cite{ottino}. The phenomenon of chaotic advection -- 
intensively investigated during the last decade -- provides a 
basic mechanism for mixing in laminar flows\cite{oreste}. Briefly 
stated, chaotic advection refers to the situation in which fluid 
elements in a non-turbulent flow follow chaotic  trajectories. 
Advection by simple time-dependent two-dimensional flows falls 
generically under this category. Stirring by chaotic motion, with 
its characteristic stretching and folding of material elements, is 
able to bring distant parts of the fluid into intimate contact and 
thus greatly enhances mixing by molecular diffusion acting at 
small scales. 

Mixing efficiency becomes specially important when the substances 
advected by the flow are not inert but have some kind of activity. 
By `activity' we mean that some time-evolution is occurring to the 
concentrations inside advected fluid elements (produced by 
chemical reactions, for example). For definiteness we will use 
terms such as chemical fields and chemical reactions, but 
biological processes, occurring for example when the advected 
substance is living plankton, can be described formally in the 
same way. The interaction between the stirring process and the 
chemical activity can result in complex patterns for the spatial 
distribution of the chemical fields, which in turn greatly affect 
the chemical processes\cite{Epstein,Edouard}. In addition to the 
impact on its own chemical dynamics, the spatial inhomogeneities 
may have important effect on other dynamical processes occurring 
in the fluid (for example in the behavior of predators seeking for 
the advected plankton \cite{Marguerit}). An understanding of the 
structure of these spatial patterns is thus valuable. 

Previous theoretical work concentrated on the temporal evolution 
of the total amount of chemical products in specific reactions 
such as $A+B \rightarrow 2C$ or $A+B \to 2B$ \cite{closedc}. In 
\cite{openc} the same type of reactions were studied in open 
flows. In a previous paper \cite{PRL} some of us considered a 
class of chemical dynamics characterized by a negative Lyapunov 
exponent in the presence of (non-homogeneous) chemical sources. 
Under such chemical processes, reactant concentrations present a 
tendency to  relax towards a local-equilibrium concentration (the 
fixed point of the local chemical dynamics). This tendency is 
disrupted by the advection process, which forces fluid elements to 
visit places with different local-equilibrium states. Depending on 
the relative strength of chaotic advection and relaxation the 
resulting concentration distribution can be smooth 
(differentiable) or exhibits characteristic filamental patterns 
that are nowhere differentiable except in the direction of 
filaments aligned with the unstable foliation induced in the fluid 
by the chaotic dynamics. The mechanism for the appearance of these 
singular filaments is similar to the one producing singular 
invariant measures in 
dynamical systems \cite{Eckmann}, although here it is affected by 
the presence of the chemical dynamics: stretching by the flow 
homogeneizes the pattern along unstable directions, whereas 
small-scale variance, cascading down from larger scales, 
accumulates along the stable directions, producing diverging 
gradients.  

The strength of the singularities of the concentration field 
$C({\bf r})$ can be characterized by a H\"older exponent $\alpha$ 
\begin{equation}
|\delta C ({\bf r_0};{\bf \delta r})| \equiv |C({\bf r_0}+{\bf 
\delta r}) - C({\bf r_0})| \sim |{\bf \delta r}|^{\alpha}, \;\;\; 
|{\bf \delta r}| \rightarrow 0 \ . 
\label{Holder}
\end{equation}
If the field is smooth (differentiable) at $\bf r_0$, $\alpha=1$, 
while for an irregular rough (e.g. filamental) structure 
$0<\alpha<1$. In \cite{PRL} we focused on the existence of a 
smooth-filamental transition as time-scales of the system are 
varied, and also obtained the most probable (bulk) value of the 
H\"older exponent. Note, however, that the H\"older exponent 
defined by (\ref{Holder}) is a local characteristic of the field, 
whose value may depend on the position ${\bf r_0}$. In this Paper 
we concentrate on such non-uniformities of the filamental chemical 
field and study how this affects scaling properties of quantities 
involving spatial averages, which are the more convenient  
quantities to be observed in experiments. 

In Section \ref{sec:transition} we review the results presented in 
\cite{PRL}, namely the smooth-filamental transition and the 
dominant value of the H\"{o}lder exponent in closed flows. Then we 
consider the same problem for the case of mixing by open flows 
(Sect.~\ref{sec:open}). In this case the necessity of a 
multifractal description becomes manifest, and this motivates the 
development of a quantitative characterization of the filamental 
structures in terms of structure functions. This is presented in 
Sect.~\ref{sec:structure}. Scaling exponents appear to be related  
to the distribution of finite-time Lyapunov exponents. We conclude 
the Paper with a summary and discussion.  

\section{The smooth-filamental transition}
\label{sec:transition}

We consider the flow as externally prescribed, thus neglecting any 
back influence of the chemical dynamics into the hydrodynamics 
(the advected substances are chemically active but 
hydrodynamically passive). In this context, the general continuum 
description of chemical reactions in hydrodynamic flows is given 
by sets of reaction-advection-diffusion equations. They involve in 
general multiple components and nonlinear reaction terms.  
Reference \cite{PRL} considered the situation in which the 
chemical kinetics is stable, i.e. there is a local-equilibrium 
state at each spatial position, determined by the sources and the 
reaction terms, so that concentrations of fluid particles visiting 
that position tend to relax to the local-equilibrium value. 
Mathematically this corresponds to the negativity of the Lyapunov 
exponents associated to the chemical dynamical subsystem. It was 
shown in \cite{PRL} that arbitrary chemical dynamics in this class 
can be substituted by linear relaxation towards local equilibrium 
at a rate given by the largest (least negative) chemical Lyapunov 
exponent. Within this restriction, the multiplicity of components 
is not essential since, except for special types of coupling, 
linearization leads to simple relations between the different 
fields. 

Because of the above remarks, and with the aim of keeping the 
mathematics as simple as possible, we will restrict our 
considerations in this Paper to the simplest chemical evolution: 
linear decay, at a rate $b$, of a single advected substance. A 
space-dependent source of the substance will also be included, to 
maintain a non-trivial concentration field at long times. This 
chemical dynamics can be considered either as an approximation to 
more complex chemical or biological evolutions, with maximum 
chemical Lyapunov exponent $-b$, or as a description of simple  
specific processes such as spontaneous decomposition of unstable 
radicals, decay of a radioactive substance, or relaxation of 
sea-surface temperature towards atmospheric values 
\cite{Abraham2}. The validity of our ideas for nonlinear 
multicomponent situations has been checked for a plankton model in 
\cite{PCE}. 

The concentration field $C({\bf r},t)$, when advected by a 
incompressible velocity field ${\bf v}({\bf r},t)$ is governed by 
the equation 
\begin{equation}
\frac{\partial C}{\partial t}+{\bf v}({\bf r},t) \cdot \nabla C= 
S({\bf r})- b C + \kappa \nabla^2 C, 
\label{Euler}
\end{equation}
where $\kappa$ is the diffusion coefficient, $b$ is the decay rate 
introduced above and $S({\bf r})$ is the concentration input from 
chemical sources (negative values representing sinks). We restrict 
our study to the case in which the incompressible velocity field 
is two-dimensional, smooth, and non-turbulent. Chaotic advection 
is obtained generically if a  simple time-dependence, for example 
periodic, is included in ${\bf v}({\bf r},t)$. We assume that 
diffusion is weak and transport is dominated by advection. Thus 
one expects that the distribution on scales larger than a certain 
diffusive scale is not affected by diffusion. Therefore we 
consider the limiting non-diffusive case $\kappa=0$. In this limit 
the above problem can be described in a Lagrangian picture by an 
ensemble of ordinary differential equations 
\begin{equation}
\frac{d \hr }{dt}={\bf v}(\hr,t),  
\label{Lagranger}
\end{equation}
\begin{equation}
\frac{d \hat C}{dt}=S[\hr(t)]-b \hat C,\; 
\label{LagrangeC}
\end{equation}
where the solution of the first equation gives the trajectory of a 
fluid parcel, $\hr(t)$, while the second one describes the 
Lagrangian chemical dynamics in this fluid element: $ \hat C(t) 
\equiv C[{\bf r}=\hr(t),t]$. 

To obtain the value of the chemical field at a selected point 
$\br$ at time $\bar t$ one needs to know the previous history of 
this fluid element, that is the trajectory $\hr(t)$ ($0 \le t \le 
\bar t$) with the property $\hr(\bar t) = \br$. This can be 
obtained by the integration of (\ref{Lagranger}) backwards in 
time. Once $\hr(t)$ has been obtained, the solution of 
(\ref{LagrangeC}) is 
\begin{equation}
C(\br,\bar t) = C[{\bf \hat r}(0),0] e^{-b \bar t} + \int_0^{\bar 
t} S[{\bf \hat r}(t)] e^{-b (\bar t-t)} dt. 
\label{field}
\end{equation}

One can obtain the difference at time $\bar t$ of the values of 
the chemical field at two different points $\br$ and $\br+\bdr$ 
separated by a small distance $\bdr$ in terms of the difference 
$C[\hr(t) + \hdr(t),t]-C[\hr(t),t] \equiv \delta 
C[\hr(t),t;\hdr(t)]$ for $0 \le t \le \bar t$, namely: 
\begin{eqnarray}
\delta C(\br,\bar t; \bdr) = \delta C[\hr(0),0;\hdr(0)] e^{-b \bar 
t}  \nonumber \\ + \int_0^{\bar t} \delta S[\hr (t); \hdr(t)] 
e^{-b (\bar t- t)} dt 
\label{deltaC1} 
\end{eqnarray}
where ${\hat {\delta {\bf r}}}(t)$ ($0 \le t \le \bar t$) is the 
time-dependent distance between the two trajectories that end at  
$\br$ and $\br+\bdr$ at time $\bar t$, and $\delta S$, in analogy 
with $\delta C$, is the difference of the source term at points 
$\hr(t)$ and $\hr(t)+\hdr(t)$. 

Thus we have expressed the behavior of an Eulerian quantity 
$\delta C(\br,\bar t; \bdr)$ in terms of Lagrangian quantities, in 
particular of $\hdr(t)$. Further analysis of Eq.~(\ref{deltaC1}) 
requires specification of the behavior of $\hdr$. The signature of 
chaotic advection is the exponentially growing behavior of this 
quantity at long times. More precise statements need additional 
assumptions on the character of the flow. The simplest framework 
is obtained if we restrict our attention to initial conditions 
$\hr(0)$ in an invariant {\sl hyperbolic} set  \cite{Eckmann}. In 
this case one can identify at each point two directions, the one 
in which the flow is {\sl contracting} ${\bf c}({\bf r})$ and the 
{\sl expanding} direction ${\bf e}({\bf r})$. They depend 
continuously on position (for time-dependent flows there is an 
additional explicit time-dependence that we do not write down to 
simplify the notation). Since ${\bf c}$ and ${\bf e}$ form at each 
point a vector basis which is not orthonormal, it is convenient to 
introduce also the dual basis (${\bf c^\dag}, {\bf e^\dag}$) at 
each point. Chaotic advection manifests in which for most initial 
separations $\hdr(0)$ the long-time behavior of $\hdr$ is $|\hdr 
(t)| \sim \left| {\bf e}[\hr(0)]^\dag \cdot \hdr(0)\right| 
e^{\lambda t}$ (for $t>0$), where $\lambda$ is the positive 
Lyapunov exponent of the flow and ${\bf e}[\hr(0)]^\dag \cdot 
\hdr(0)$ gives the component of the initial separation along the 
expanding direction. At long times, the direction of $\hdr (t)$ 
tends to become aligned with the expanding direction of the flow 
at $\hr(t)$, ${\bf e}[\hr(t)]$. However, if the initial separation 
is aligned with the contracting direction at the initial point, 
$\lambda$ should be substituted by $\lambda'$, the contractive 
Lyapunov exponent, and ${\bf e}^\dag$ by ${\bf c}^\dag$. For 
incompressible flows one has $\lambda'=-\lambda$. 

In order to  analyze (\ref{deltaC1}) one has to consider the 
backwards evolution. In this case typical solutions behave, for 
$t<0$ and large, as $|\hdr (t)| \sim \left| {\bf 
c}[\hr(0)]^\dag\cdot\hdr(0) \right| e^{\lambda' t} = \left| {\bf 
c}[\hr(0)]^\dag\cdot\hdr(0) \right| e^{-\lambda t}$ so that, also 
in this backwards dynamics, close initial conditions diverge, and 
the difference will tend to become aligned with the most expanding 
direction of the backwards flow (the contracting of the forward 
flow, ${\bf c}[\hr(t)]$). Again, there is a particular direction 
for the orientation of the initial condition (the contracting one  
in the backwards flow which is the expanding one in the forward 
dynamics) for which $-\lambda$ should be replaced by $\lambda$. In 
(\ref{deltaC1}), $\hdr$ is obtained backwards starting from $\bdr$ 
at $t=\bar t$. In this case: 
\begin{equation}
\hdr(t) \approx  {\bf c(\br)}^\dag \cdot\bdr \ e^{\lambda(\bar t 
-t)}{\bf c}[\hr(t)] \ ,\;\; t_s \le t \le\bar t \ .
\label{dispersion}
\end{equation}
The exponential separation (\ref{dispersion}) holds only while the 
distance $\hdr(t)$ is not too large, and saturates when 
approaching the size of the system or some characteristic 
coherence length of the velocity field. The time at which this 
happens defines $t_s$, a saturation time. We assume that both the 
velocity field and the source $S({\bf r})$ have only large-scale 
structures such that their corresponding coherence lengths are 
comparable to the system size, that we take as the unit of 
length-scales. Thus, the saturation time is given by $t_s= \bar t+ 
\lambda^{-1} \ln |\bdr|$. 

For small $\bdr$, Eq.~(\ref{deltaC1}) can be written as 
\begin{eqnarray}
\delta C(\br,\bar t; \bdr) \approx \delta C[\hr(0),0;\hdr(0)] 
e^{-b \bar t}  \nonumber \\ + \int_0^{t_s} \delta S[\hr (t); 
\hdr(t)] e^{-b (\bar t- t)} dt \nonumber \\ + \int_{t_s}^{\bar t} 
\hdr(t)\cdot \nabla S[\hr (t)] e^{-b (\bar t- t)} dt 
\label{deltaC}
\end{eqnarray}
(if $t_s>0$). We will not need to specify the dependence of 
$\delta S$ on $\hdr$ for times previous to $t_s$, as long as 
$\delta S$ remains bounded. Substitution of (\ref{dispersion}) in 
the second integral leads to 
\begin{eqnarray}
\delta C(\br,\bar t; \bdr) \approx \delta C[\hr(0),0;\hdr(0)] 
e^{-b \bar t}  \nonumber \\ + \int_0^{t_s} \delta S[\hr (t); 
\hdr(t)] e^{-b (\bar t- t)} dt \nonumber 
\\ 
 +  \bdr \cdot {\bf c(\br)}^\dag \int_{t_s}^{\bar t}  
{\bf c}[\hr(t)]\cdot \nabla S [\hr(t)] e^{(\lambda -b)(\bar t-t)} 
dt 
\label{deltaCfinal}
\end{eqnarray}

Taking the limit $\bar {\bf \delta r} \rightarrow 0$ (for a finite 
$\bar t$) leads to $t_s<0$. Thus the first integral disappears and 
the first term can be linearized. By writing $\bdr= \bar {\bf n} 
|\bdr|$ so that $\bar {\bf n}$ is a unit vector, one finds the 
directional derivative along the direction of $\bar {\bf n}$ as 
\begin{eqnarray} 
\bar {\bf n} \cdot \nabla C(\br,\bar t)\approx \bar {\bf n} \cdot 
{\bf c(\br)}^\dag {\bf c}[\hr(0)]\cdot \nabla C[\hr(0),0] 
e^{(\lambda-b)\bar t}  
 \nonumber 
\\  +  \bar {\bf n} \cdot {\bf 
c(\br)}^\dag \int_{0}^{\bar t} {\bf c}[\hr(t)]\cdot \nabla S 
[\hr(t)] e^{(\lambda -b)(\bar t-t)} dt  \ .
\label{directional}
\end{eqnarray}

If $\lambda < b$ this derivative remains finite in the $\bar t 
\rightarrow \infty$ limit and the asymptotic field $C_\infty(\br) 
\equiv C({\bf r},\bar t \rightarrow \infty)$ is smooth 
(differentiable). Otherwise the derivatives of $C$ diverge as 
$\sim e^{(\lambda -b)\bar t}$ leading to a nowhere-differentiable 
irregular asymptotic field. The exception again is the expanding 
direction of the forward flow: when $\bdr$ points along that 
direction one should substitute in Eq.~(\ref{directional}) 
$\lambda$ and ${\bf c}$ by $-\lambda$ and ${\bf e}$. This 
directional derivative is always finite. Thus, there is at each 
point a direction along which $C_\infty$ is smooth. By changing 
the values of $b$ and $\lambda$ one will encounter, at $b=\lambda$ 
a morphological transition between a smooth pattern and a 
filamental (i.e., singular in all but one directions) one. It 
should be noted that, because the explicit time-dependence of the 
vectors ${\bf c}$ and ${\bf e}$ referred to before, the limiting 
distribution $C_\infty$ will not be a steady field, but one 
following the time dependence of the stable and unstable 
directions. For time-periodic flows ${\bf v}({\bf r},t)$, 
$C_\infty$ will also be time periodic. Its singular 
characteristics however do not change in time. 

In order to characterize the singular asymptotic field we take the 
limit $\bar t \rightarrow \infty$ for fixed finite $\bar {\bf 
\delta r}$ in (\ref{deltaCfinal}) 
\begin{eqnarray}
\delta C_\infty (\br;\bdr) \approx 
\int_{|\bdr|^{-1/\lambda}}^{\infty} \delta S[\hr (x); \hdr(x)] 
x^{-b-1} dx 
 \nonumber
\\  +  |\bdr| \bar {\bf n} \cdot {\bf
c(\br)}^\dag  \int_{1}^{|\bdr|^{-1/\lambda}} {\bf c}[\hr(x)]\cdot 
\nabla S[\hr(x)] x^{\lambda-b-1} dx 
\end{eqnarray}
where we used the change of variables $e^{\bar t-t} \rightarrow 
x$. If $b> \lambda$ one finds for the dominant term in the $|\bar 
{\bf \delta r}| \rightarrow 0$ limit the simple scaling $\delta 
C_\infty \sim |\bdr|$, but when $b<\lambda$:  
\begin{equation}
\delta C_\infty (\br;\bdr) \sim |\bdr|^{\frac{b}{\lambda}} \ . 
\label{Holder1}
\end{equation}
According to (\ref{Holder}) the value of the H\"older exponent is 
\begin{equation}
\alpha = \min \left\{ \frac{b}{\lambda} , 1 \right\}. 
\label{Holder2} 
\end{equation} 
This implies that if $b<\lambda$ the asymptotic chemical field 
becomes an irregular fractal object (again there is an orientation 
of $\bdr$ along which $\alpha=1$). Consequently, the graph of the 
field along a one-dimensional cut or transect, or the contours of 
constant concentration are also fractals, as they are 
two-dimensional sections of the whole $C_\infty(\bf r)$ embedded 
in a three-dimensional space. More precisely, the 
one-dimensional transect (along the direction $x$) of the field is 
a self-affine function with its graph embedded in an inherently 
anisotropic space ($C-x$) with axis representing different 
physical quantities. Contours of constant concentrations, however, 
are self-similar fractal sets of the two-dimensional physical 
space ($x-y$). 

\begin{figure}
\epsfig{file=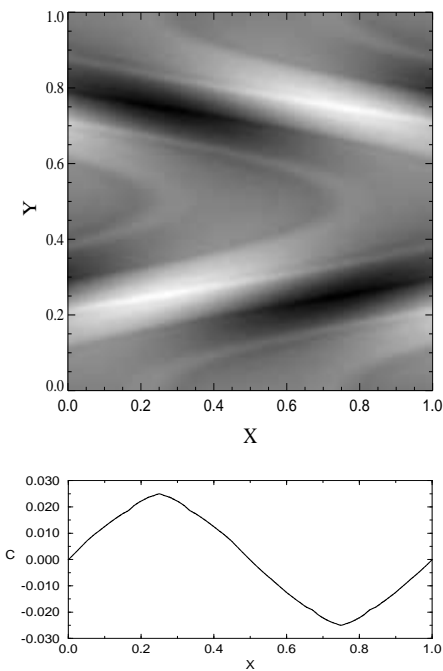,width=.8\linewidth} \caption{Top: a 
snapshot of the chemical concentration $C_\infty$ obtained for 
$b=4.0$ and $\lambda=2.67$, so that a smooth distribution is 
obtained. Bottom: a horizontal cut along the line $y=0.25$.} 
\label{fig:closedS}
\end{figure}

\begin{figure}
\epsfig{file=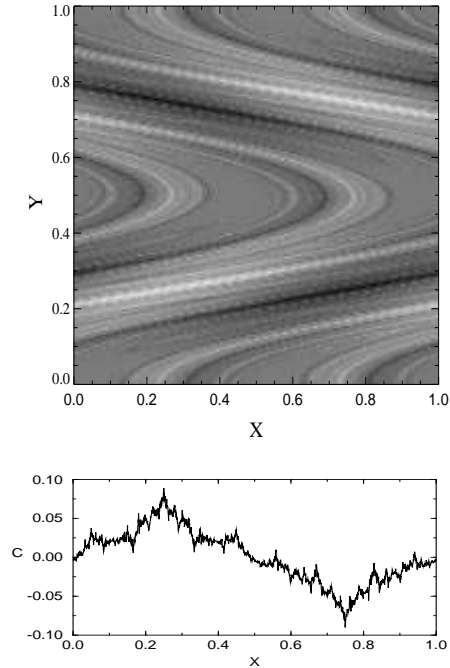,width=.8\linewidth} \caption{Top: a 
snapshot of the chemical concentration $C_\infty$ for $b=0.1$ and 
$\lambda=2.67$, so that a filamental structure is obtained. The 
lower panel shows a horizontal cut along the line $y=0.25$, 
clearly displaying the fractal nature of the field.} 
\label{fig:closedF}
\end{figure}

In Figs.~\ref{fig:closedS} and \ref{fig:closedF} we present 
snapshots of the asymptotic field $C_\infty$ evolving according to 
(\ref{Lagranger})-(\ref{LagrangeC}). For the flow we take a simple 
time-periodic velocity field defined in the unit square with 
periodic boundary conditions by 
\begin{eqnarray}
v_x(x,y,t)& =& -\frac{2U}{T} \Theta \Bigl(\frac{T}{2}-t \bmod T 
\Bigr) \cos({2\pi y}) \nonumber \\ v_y(x,y,t)&=& -\frac{2U}{T} 
\Theta \Bigl(t \bmod T-\frac{T}{2} \Bigr) \cos({2\pi x}) 
\label{flow}
\end{eqnarray}
where $\Theta(x)$ is the Heavyside step function. In our 
simulations $U=1.2$, which produces a flow with a single connected 
chaotic region. The value of the numerically obtained Lyapunov 
exponent is $\lambda \approx 2.67/T$. 

Backward trajectories with initial coordinates on a rectangular 
grid were calculated and used to obtain the chemical field at each 
point by using (\ref{field}) forward in time with the source term 
$S(x,y)=1+\sin{(2 \pi x)}\sin{(2 \pi y)}$. The values of the 
parameters used in Fig.~\ref{fig:closedS} are $T=1.0$ and $b=4.0$, 
for which the Lyapunov exponent is $\lambda=2.67<b$. A smooth 
pattern is seen, in agreement with our theoretical arguments. In  
Fig.~\ref{fig:closedF} the parameters are $T=1.0$ and $b=0.1$, so 
that $\lambda>b$ and a filamental pattern is obtained. 

The smooth-fractal transition also appears in the time-dependence 
of the concentration measured at a fixed point in space. This can 
be shown by a similar analysis for the difference 
\begin{equation}
\delta C({\bf r},t;\delta t) \equiv C({\bf r},t+\delta t)-C({\bf 
r},t) 
\end{equation}
instead of the spatial difference discussed above. If $b<\lambda$ 
the signal $C({\bf r},t)$ becomes non-differentiable in time and 
can be characterized by the same H\"older exponent $b / \lambda$. 
The fact that scaling properties of the temporal signal and that 
of the spatial structure are the same --analogously to the so 
called `Taylor hypothesis' in turbulence-- can be exploited in 
experiments or in analysis of geophysical data. 

We conclude with some comments on the range of validity of 
Eq.~(\ref{directional}). The Lagrangian description 
(\ref{Lagranger})-(\ref{LagrangeC}) in which our approach is based 
is valid only for scales at which diffusion is negligible. Thus 
there is a minimum admissible value $l_{diff}(\kappa) \sim 
\sqrt{\kappa}$ of $\bdr$ and our calculation should be understood 
as giving the gradients only up to this scale, fractality being 
washed out at smaller scales by the presence of diffusion. 
Nevertheless we think that, if diffusion is weak, the 
fractal-filamental transition will be seen at scales larger than 
this diffusive scale. For fixed $\bdr$ larger than the diffusion 
length, (\ref{deltaCfinal}) remains valid until a time $\bar t 
\lesssim t_s(\bdr)$ that means that the divergence of the 
gradients will also saturate at a finite value   $\sim 
(l_{diff})^{b/\lambda -1}$. 

Another limitation to the validity of our equations arises from 
the fact that, for most chaotic flows of physical relevance, not 
all the points visited by the fluid particle will be hyperbolic. 
The stable and unstable directions in the previous discussion  
become tangent at some points and equations such as 
(\ref{deltaCfinal}) become undefined there. More importantly, KAM 
tori will be present in the system, so that for values of $\br$ 
lying on KAM trajectories the value of the Lyapunov exponent 
appearing in (\ref{dispersion}) will not be positive but zero.  
These facts imply that our previous equations are not valid in a 
global sense. The main effect of the KAM tori will be to partition 
space into ergodic regions. Each region will be characterized by a 
different value of $\lambda$, and there is the possibility of 
observing different morphologies in the different regions. For the 
values of parameters used in Figs.~\ref{fig:closedS} and 
\ref{fig:closedF}, KAM tori occupy a very small and practically 
unobservable portion of space, so that the filamental pattern 
appears to be well described by the same H\"{o}lder exponent 
nearly everywhere. However, an example is given in \cite{PRL} in 
which filamental and smooth regions coexist separated by KAM tori. 

It is known that, even inside ergodic regions, the Lyapunov 
exponent is a non-smooth function of space\cite{Ott}: it can 
differ from the most probable value $\lambda$ in sets of zero 
Lebesgue measure. This stresses again that our arguments above 
have a local nature. Equations such as (\ref{Holder2}) give the 
value of the `bulk' or `most probable' H\"{o}lder exponent, that 
is, the one which characterizes almost all points in the fluid. 
More generally one should define local H\"{o}lder exponents 
$\alpha(\br)$ which depend on space. Deviations from the most 
probable or bulk value occur only in portions of the ergodic 
regions having vanishing measure. They can however have effects on 
the physically observable average quantities. These effects appear 
in a specially clear manner when discussing open flows, the 
subject of the next Section.  

\section{Open flows}
\label{sec:open}

Let us now consider again the problem 
(\ref{Lagranger})-(\ref{LagrangeC}) with a velocity field 
corresponding to an open flow whose time-dependence is restricted 
to a finite region (mixing region), with asymptotically steady 
inflow and outflow regions. A prototype of this flow structure is 
a stream passing around a cylindrical body. If the inflow velocity 
is high enough vortices formed in the wake of the cylinder make 
the flow time-dependent in this region, while the flow remains 
steady in front of the cylinder or in the far downstream region. 
We assume again that the flow is non-turbulent, so that the 
velocity field is spatially-smooth. 

Passive advection in such open flows was found to be a nice 
example of chaotic scattering\cite{Tel,Zemniak}. Advected 
particles (or fluid elements) enter the unsteady region, undergo 
transient chaotic advection, and finally escape and move away 
downstream on simple orbits. The time spent in the mixing region, 
however, depends strongly on the initial coordinates, with 
singularities on a fractal set corresponding to particles trapped 
forever in the mixing region. This is due to the existence of a 
non-attracting chaotic set (although of zero measure) formed by an 
infinite number of bounded hyperbolic orbits in the mixing region. 
The stable manifold of this chaotic set (or chaotic saddle) 
contains orbits coming from the inflow region but never escaping 
from the mixing zone. These points correspond to the singularities 
of the residence time. If a  droplet of dye is injected into the 
mixing region, most of it will be advected downstream in a short 
time. But part of the dye will remain close to the  chaotic saddle 
for very long times, and continuously ejected along its unstable 
manifold. In this way the dye traces out the unstable manifold of 
the chaotic saddle, resulting in fractal patterns characteristic 
to open flows\cite{Jung,Sommerer}. Permanent chaotic advection is 
restricted to a fractal set of zero Lebesgue measure. Points close 
to the unstable manifold of the chaotic saddle have spent a long 
time in the mixing region of the flow moving near chaotic orbits 
with a positive Lyapunov exponent. For points precisely at this 
unstable manifold, the backwards trajectories (the ones from which 
the Lyapunov exponent in (\ref{Holder2}) should be computed)  
remain in the chaotic saddle, thus leading to $\lambda>0$. The 
other trajectories escape from the chaotic set in a short time, 
thus being characterized by a Lyapunov exponent equal to zero. 

Thus open flows provide a rather clear example of strong 
space-dependence 
of Lyapunov exponents. According to Eq.~(\ref{Holder2}), the 
H\"{o}lder exponent may be different from $1$ only on the unstable 
manifold of the chaotic saddle, thus implying that the transition 
from smooth to filamental structure now only takes place in this 
fractal set of zero measure. The background chemical field is 
always smooth, independently of the value of $b$. 

\begin{figure}
\epsfig{file=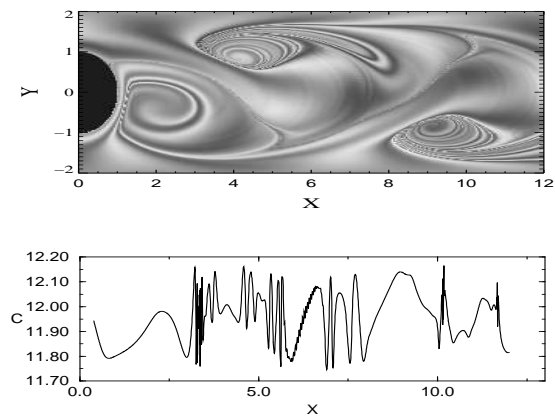,width=.95\linewidth } \caption{Top: a 
snapshot of the chemical pattern $C_\infty$ formed in the wake of 
a cylinder (the black semicircle at the left is half of its 
section). Mean flow is from left to right. We have used the 
streamfunction given in reference [15]. $b=0.96$, and all the 
streamfunction parameter values are the same as in [15] except the 
boundary-layer thickness of the cylinder which here takes the 
value $a=20.0$, and the vorticity strength which is $\omega=35.06$ 
in our calculations. Bottom: a horizontal cut taken along the line 
$y=1.0$.} 
\label{fig:openflowC} 
\end{figure}

\begin{figure}
\epsfig{file=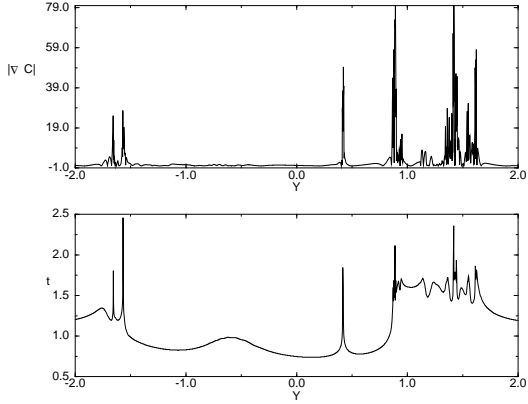,width=.9\linewidth} \caption{Top: The 
absolute value of the gradient of the chemical field in 
Fig.~\protect{\ref{fig:openflowC}} along the line $x=7.3$. The 
lower figure shows the escape time for particles along this line 
($x=7.3$). This is calculated by computing the time that every 
single particle takes to arrive to the line $x=-2.0$ (far from the 
chaotic wake region) in the backwards-in-time dynamics.} 
\label{fig:openflowD}
\end{figure}

To check these ideas, we obtain numerically the distributions of 
chemical fields advected by an open flow. Our velocity field is 
taken from a kinematic model of a time-periodic flow behind a 
cylinder, described in \cite{Jung}. This flow was found to be 
qualitatively similar to the solution of the Navier-Stokes 
equation in the range of the Reynolds number corresponding to 
time-periodic vortex seeding. The concentration pattern shows 
irregularities separated by smooth regions 
(Fig.~\ref{fig:openflowC}, obtained with $b=0.96$). This is more 
clearly observed in the longitudinal transect. The relation 
between the singular regions and the location of the chaotic 
saddle can be made patent by comparing the gradient of the field 
with the spatial dependence of the escape times from the 
scattering region. In particular, Fig.~\ref{fig:openflowD} shows 
the absolute value of the gradient of the concentration field 
$|\nabla C_\infty({\bf r})|$ that is highly intermittent. It also 
displays the time (in the time-reversed dynamics) that fluid 
particles initially in a line perpendicular to the mean flow take 
to escape the region of chaotic motion. Most of the particles 
leave the region in a short time, but longer times appear for 
initial locations close to the chaotic saddle. Clearly, these 
diverging times are associated to the singularities in the 
gradient distribution. By increasing the value of $b$ the flow 
characteristics (trajectories, manifolds, escape times, ...) 
remain unchanged, but the singularities in the advected field 
decrease and finally a smooth distribution is obtained.  

A chemical field with the same structure can also be obtained in 
open flows whose time-dependence (and thus the chaoticity of 
advection) is not restricted to a finite domain, by restricting 
the spatial dependence of the chemical sources to a finite region. 
This case was investigated in the context of plankton dynamics in 
\cite{PCE}. 

Since the irregularities now appear only on a set of measure zero, 
one could ask if they can have any significant effect on 
measurable quantities. In order to clarify this, instead of the 
previous characterization of the point-wise strength of the 
singularities by the H\"older exponent, let us investigate the 
scaling of the spatial average of the differences $\delta 
C_\infty$ with distance $\delta r \equiv |\delta{\bf r}|$. On a 
one-dimensional transect of unit length the total number of 
segments of length $\delta r$ is $(\delta r)^{-1}$ while the 
number of segments containing parts of the unstable manifold (with 
partial fractal dimension $\tilde D$) is $N(\delta r) \sim {\delta 
r}^{-\tilde D}$. Thus, according to (\ref{Holder2}) the spatial 
average of $\delta C_\infty$ along this line, $\langle \delta 
C_\infty({\bf r};\delta r) \rangle$, can be written as 
\begin{eqnarray}
\langle \delta C_\infty({\bf r};\delta r) \rangle = (\delta r) 
(\delta r)^{-\tilde D} (\delta r)^{b/\lambda} + \nonumber \\ + 
(\delta r) [ (\delta r)^{-1} - (\delta r)^{-\tilde D} ] (\delta r) 
\end{eqnarray}
where the first term pertains to the singular, while the second 
one to the smooth component. In the limit $\delta r \rightarrow 0$ 
the dominating behavior is 
\begin{equation}
\langle \delta C_\infty({\bf r};\delta r) \rangle \sim \delta 
r^\zeta\ \ , 
\end{equation}
with 
\begin{equation}
\zeta = \min\left\{ 1, 1 + {b \over \lambda} - \tilde D \right\} 
\label{zeta}
\end{equation}
showing that if $\tilde D<{b \over \lambda}$ the average is 
dominated by the smooth component, but if the singularities are 
strong enough (or the fractal dimension of the singular set is 
large enough) they contribute to the scaling of $\langle \delta 
C_\infty({\bf r};\delta r) \rangle$. The average has been 
performed along a one-dimensional line or transect of the 
two-dimensional pattern. For common velocity fields and transects 
this will be equivalent to the complete average over the whole 
fluid, except in the particular case in which the transect is 
chosen to be completely aligned with the filaments. 

\section{Structure functions}
\label{sec:structure}

The strongly intermittent structure of singularities in open flows 
is an extreme example. There are additional inhomogeneities 
affecting both to the open and to the closed flows: although, in 
the long-time limit the Lyapunov exponent is the same for almost 
all trajectories in an ergodic region, deviations can persist on 
fractal sets of measure zero, and as we saw above such sets can 
contribute significantly to the global scaling. The origin of 
these inhomogeneities can be traced back by analyzing the 
finite-time distribution of Lyapunov exponents. This will be done 
in the following. For a robust quantitative characterization of 
the filamental structure, accessible to measurements, we consider 
now the scaling properties of the structure functions associated 
with the chemical field. 

The $q$th order structure function is defined as 
\begin{equation}
S_q(\delta r) = \langle |\delta C_\infty({\bf r};\delta r)|^q 
\rangle 
\label{Sq}
\end{equation}
where $\langle\ \rangle$ represents averaging over different 
locations ${\bf r}$, and $q$ is a parameter (we will only consider 
structure functions of positive order ($q>0$)). In the absence of 
any characteristic length over a certain range of scales the 
structure functions are expected to exhibit a power-law dependence 
\begin{equation}
S_q(\delta r) \sim \delta r^{\zeta_q} 
\label{zetaq}
\end{equation}
characterized by the set of scaling exponents $\zeta_q$. 

We also note that some of the scaling exponents are directly 
related to other characteristic exponents, such as the one 
characterizing the decay of the Fourier power-spectrum $\Gamma(k) 
\sim k^{-\gamma}$, or the box-counting fractal dimension $D_G$ of 
the graph of the function $C_\infty(x,y)$ as a function of $x$ by 
simple relations \cite{Bohr}: 
\begin{equation}
\gamma = \zeta_2 +1 \;\;\; {\rm and} \;\;\;  D_G = 2 - \zeta_1. 
\label{otherexponents}
\end{equation}
If the H\"older exponent of the field has the same value 
everywhere, given by (\ref{Holder2}), the scaling exponents of the 
resulting {\it mono-affine} field are simply 
\begin{equation}
\zeta_q= q \alpha = q {b \over \lambda}. 
\end{equation}
(we have assumed $b<\lambda$). In the case of our chemical field 
this, however, would only hold in the non-generic case in which 
the stretching by the flow is spatially uniform (e.g. like in a 
simple area-preserving baker's map). In general, the singular 
spatial inhomogeneities of the Lyapunov exponent could be 
understood by realizing that the finite-time stretching rates, or 
finite-time Lyapunov exponents \cite{Ott}, have a certain 
distribution around the most probable value. This distribution 
approaches the time-asymptotic form \cite{Ott,Bohr}: 
\begin{equation}
P(\lambda, t) \sim t^{1/2} e^{-G(\lambda)t} 
\label{finitelyap}
\end{equation}
where $G(\lambda)$ is a function characteristic to the system 
(velocity field, in this case), with the property that 
$G(\lambda_0)=G'(\lambda_0)=0$ and $G(\lambda)>0$, where 
$\lambda_0$ is the most probable value of the infinite-time 
Lyapunov exponent. The form (\ref{finitelyap}) is valid only for 
hyperbolic systems. Non-hyperbolicity can strongly affect the 
distribution at small values of $\lambda$ but around $\lambda_0$ 
and for larger values it remains a good approximation. As we shall 
see later only this region contributes to the structure functions 
of positive order. 

 As time increases the distribution 
becomes more and more peaked around $\lambda_0$. 
The (Lebesgue) area of the set of initial conditions with 
finite-time Lyapunov exponents in a small interval $(\lambda, 
\lambda+\delta\lambda)$ that excludes $\lambda_0$ decreases at 
long times with a dominant exponential behavior:  
\begin{equation}
\delta A_{\lambda}(t) \sim e^{-G(\lambda)t}\delta\lambda 
\label{area}
\end{equation}
showing that only sets of measure zero can have Lyapunov exponent 
different from $\lambda_0$ in the $t \rightarrow \infty$ limit. 
Such sets, however, can still have nonzero fractal dimensions. At 
finite times, the area (\ref{area}) encloses the final anomalous 
set, with a transverse thickness that, due to stretching by the 
chaotic advection, decreases like $l_\lambda (t) \sim e^{-\lambda 
t}$. The number of boxes needed to cover the set of area $\delta 
A_{\lambda}(t)$ using boxes of size $l_{\lambda} (t)$ is 
\begin{equation}
N_{\lambda}(t) \sim {\delta A_\lambda(t) \over l^2(t)} \sim e^{[2 
\lambda - G(\lambda)]t} \sim l(t)^{{G(\lambda) \over \lambda}-2} 
\end{equation}
that gives the dimension for the set to which this area converges 
in the infinite-time limit:  
\begin{equation}
D(\lambda) = 2 - {G(\lambda) \over \lambda}. 
\end{equation}
Thus, an arbitrary line across the system will be found composed  
by subsets of dimension $\tilde D(\lambda)=D(\lambda)-1$ each one 
characterized by different values $\lambda$ of the Lyapunov 
exponents and in consequence of the H\"older exponents 
$\alpha(\lambda) = \min\{ b/\lambda$, 1\}. 

Now, the scaling exponents in (\ref{zetaq}) can be readily 
obtained. The number of segments of size $\delta r$ belonging to a 
subset characterized by Lyapunov exponent $\lambda$ scales as  
$N(\lambda) \sim \delta r^{-\tilde D(\lambda)}$, while the total 
number of such non-overlapping segments scales as $\sim \delta 
r^{-1}$. Thus, the structure function can be written as 
\begin{eqnarray}
S_q(\delta r) \sim \int_{\lambda_{min}}^{\lambda_{max}} \delta 
r^{2-D(\lambda)} |\delta C_\infty(r(\lambda),\delta r)|^q d\lambda 
\sim \nonumber \\ \sim \int_{\lambda_{min}}^{b} \delta 
r^{2-D(\lambda)} \delta r^{ q} d\lambda + \int_{b}^{\lambda_{max}} 
\delta r^{2-D(\lambda)} \delta r^{ q b/\lambda} d\lambda 
\end{eqnarray}
In the limit $\delta r \rightarrow 0$ the integrals are dominated 
by a saddle point and, after some manipulations, the scaling 
exponents in (\ref{zetaq}) are obtained as 
\begin{equation}
\zeta_q = \min_{\lambda} \left \{q, {q b \over \lambda} +2- 
D(\lambda) \right \} = \min_{\lambda} \left \{q, {q b+G(\lambda) 
\over \lambda} \right \} 
\label{scalingexp}
\end{equation}
The right hand side can be seen as a family of lines in the 
$q-\zeta_q$ plane labeled by the parameter $\lambda$ so that the 
value of $\zeta_q$ is given by the lower envelope of these lines. 
Note that the shape of $G(\lambda)$ for $\lambda$ small enough 
becomes irrelevant for determining $\zeta_q$ because of the 
minimum condition. Thus multifractality, characterized by 
nonlinearity in the $q$-depencence of $\zeta_q$, is affected only 
by the largest stretching rates in the flow. Equation (\ref{zeta}) 
is a particular case of (\ref{scalingexp}) for $q=1$ and in the 
approximation of considering a single value of $\lambda$ on the 
chaotic saddle.  

According to (\ref{scalingexp}) the $q$th order structure function 
is dominated by a subset characterized by a Lyapunov exponent 
$\lambda_q$. Applying the extremum condition to (\ref{scalingexp}) 
we obtain an equation for $\lambda_q$ 
\begin{equation}
{d \over d\lambda} G(\lambda)|_{\lambda = \lambda_q} = {q b + 
G(\lambda_q) \over \lambda_q} 
\label{lambdaq}
\end{equation}
that can be substituted into (\ref{scalingexp}) to obtain the 
$q$th order scaling exponent. 

\begin{figure}
\epsfig{file=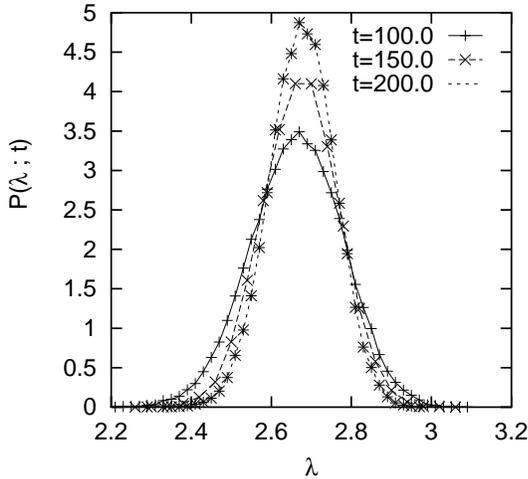,width=.8\linewidth} \caption{The distribution  
of finite-time Lyapunov exponents at three different times, 
obtained for the closed flow (\protect{\ref{flow}}) with $T=1.0$ 
and $U=1.2$. The long-time mean Lyapunov exponent is 
$\lambda_0=2.67$. } 
\label{fig:FTLE}
\end{figure}

\begin{figure}
\epsfig{file=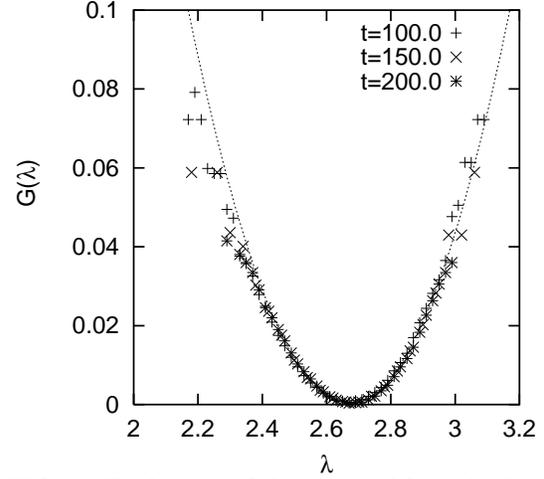,width=.8\linewidth} \caption{The function 
$G(\lambda)$, obtained from the distributions in 
Fig.~\protect{\ref{fig:FTLE}} and 
Eq.~\protect{(\ref{finitelyap})}. Collapse of data for the three 
times into the same curve confirms the validity of 
\protect{(\ref{finitelyap})}. The dotted line is a parabolic fit 
($0.4 (\lambda-2.67)^2$) that provides a good approximation near 
the minimum. } 
\label{fig:Glambda}
\end{figure}

We have analyzed numerically the chemical decay under advection by 
the closed flow (\ref{flow}) to check the theoretical predictions 
above. Numerically computed histograms of the finite-time Lyapunov 
exponents are shown in Fig.~\ref{fig:FTLE} and the corresponding 
$G(\lambda)$ functions are represented in Fig.~\ref{fig:Glambda}. 
The $G(\lambda)$ functions obtained from histograms corresponding 
to different times coalesce except in the small-$\lambda$ region, 
thus confirming that (\ref{finitelyap}) correctly describes the 
observed distribution. 

\begin{figure}
\epsfig{file=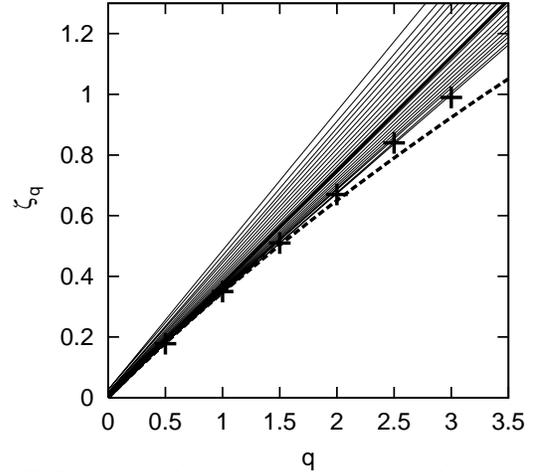,width=.8\linewidth} \caption{The scaling 
exponents $\zeta_q$ for the situation of 
Figs.~\protect{\ref{fig:FTLE}} and \protect{\ref{fig:Glambda}} and 
with $b=1.0$. Thick line: the mono-fractal approximation 
$\zeta_q=q b/\lambda_0$. Thin lines: the curves $\zeta_q=q$ and 
$\zeta_q=[qb+G(\lambda)]/\lambda$, for different values of 
$\lambda$; the numerical values of $G(\lambda)$ are obtained from 
Fig.~\protect{\ref{fig:Glambda}}. According to 
Eq.~(\protect{\ref{scalingexp}}), the actual values of the scaling 
exponents are given by the lower envelope of this set of curves. 
This is confirmed by the numerically determined values of 
$\zeta_q$ (crosses). Dashed line: the approximation 
(\protect{\ref{Chertkovscaling}}).  } 
\label{fig:scaling}
\end{figure}

Numerically calculated scaling exponents (i.e. obtained by direct 
application of Eq.~(\ref{Sq}) averaging over many one-dimensional 
transects), and the family of lines corresponding to 
(\ref{scalingexp}) based on the histogram of the finite-time 
Lyapunov exponents of Fig.~\ref{fig:FTLE} are shown in 
Fig~\ref{fig:scaling}, where the prediction of the mono-fractal 
approximation ($\zeta_q = q b/\lambda_0$) is also shown. The 
mono-fractal approximation appears to be accurate for small $q$. 
The graph-fractal dimension or the widely used Fourier power 
spectrum exponent are related to $\zeta_1$ and $\zeta_2$ by 
Eqs.~(\ref{otherexponents}) so that their estimate based on the 
mono-fractal description that considers just the bulk value of the 
H\"older exponent can deviate from the actual values. 

In a recent work by Nam et al. \cite{Nam} the power spectrum of a 
decaying scalar field (with space-dependent decay rate) has been 
investigated and related to the distribution of finite-time 
Lyapunov exponents of the advecting flow. The result for the 
spectral exponent obtained in \cite{Nam} using an eikonal-type 
wave packet model \cite{Antonsen}, and taking into account finite 
diffusion, is consistent with our formula (\ref{scalingexp}) (that 
for $q=2$, and with (\ref{otherexponents}) gives the value of the 
spectral slope) obtained in the non-diffusive limit.

The function $G(\lambda)$ is characteristic to the advecting flow. 
Let us now consider a special case where we approximate 
$G(\lambda)$ by a parabola 
\begin{equation}
G(\lambda) = {(\lambda - \lambda_0)^2 \over 2 \Delta} \ . 
\label{Gauss}
\end{equation}
This can be thought as the first term in a Taylor expansion around 
$\lambda_0$, which is a good approximation to obtain the small-$q$  
scaling exponents. In this case (\ref{lambdaq}) can be solved 
explicitly 
\begin{equation}
\lambda_q = \sqrt{ (\lambda_0)^2 + 2 q b \Delta } \ . 
\end{equation}
This gives the scaling exponents 
\begin{equation}
\zeta_q = \sqrt{ \left ({\lambda_0 \over \Delta} \right )^2 + {2qb 
\over \Delta}} - {\lambda_0 \over \Delta}. 
\label{Chertkovscaling}
\end{equation}

The above relation has been obtained recently by Chertkov in  
\cite{Chertkov}, where the problem of advection of decaying 
substances was considered in a probabilistic set-up, using 
stochastic chemical sources and a random velocity field that is 
spatially smooth but delta-correlated in time. The distribution of 
stretching rates was assumed to be Gaussian as in (\ref{Gauss}). 
This assumption could be realistic in many cases, and could give 
good estimates for the scaling exponents for small $q$. For 
higher-order moments, however, higher-order terms in the expansion 
of $G(\lambda)$ can become important. Moreover, the possible 
values of $\lambda$ could be limited by a finite maximum value 
$\lambda_{max}$, e.g. in time-periodic flows, where the finite 
time-Lyapunov exponents cannot have arbitrarily large values. This 
implies that the scaling exponents for $q>q^*$, where 
$\lambda_{q^*} = \lambda_{max}$, should display a simple linear 
dependence 
\begin{equation}
\zeta_q = { qb+G(\lambda_{max}) \over \lambda_{max}} 
\end{equation}
that differs from the $q^{\frac{1}{2}}$ behavior of 
(\ref{Chertkovscaling}) for large $q$.

\section{Summary and discussion}
\label{sec:summary}

Multifractality of advected fields generated by chaotic advection 
has been observed previously in the case of passive advection with 
no chemical activity ($b=0$) \cite{OttAntonsen}. It was shown that 
the measure defined by the gradients of the advected scalar field 
has multifractal properties and its spectra of dimensions $D_q$ 
has been related to the distribution of finite-time Lyapunov 
exponents \cite{OttAntonsen}. This multifractality, however, does 
not affect the slope of the power spectrum 
\cite{AntonsenOtt,PRLcomment} (the so called Batchelor spectrum: 
$\Gamma(k) \sim k^{-1}$) or scaling exponents of the structure 
functions ($\zeta_q = 0$ for all $q$, as it can be seen from 
(\ref{scalingexp}) by setting $b=0$). The effect of 
multifractality on the power spectrum has a character transient in 
time, moving towards smaller and smaller scales and finally 
disappearing when reaching the diffusive end of the spectrum. In 
the stationary state only the diffusive cut-off of the power 
spectrum is affected \cite{Yuan} that can still be important for 
the interpretation of some experimental results. By comparing 
these results for the conserved case with the ones presented here 
for the decaying scalar we can conclude that, although the origin 
of the multifractality is the same in both situations --the 
non-uniformity of the finite-time Lyapunov exponents-- in the 
presence of chemical activity this has stronger consequences 
(non-Batchelor power spectra and anomalous scaling). 

We have presented a simple mechanism that can generate 
multifractal (or more precisely, multi-affine) distributions of 
advected chemical fields. The main ingredients are chaotic 
advection and linear decay of the advected quantity in the 
presence of non-homogeneous sources. Essentially the same 
mechanisms were considered in \cite{Chertkov}, but with stochastic 
time-dependencies both in the flow as in the chemical sources, 
considering advection by the spatially smooth limit of the 
so-called Kraichnan model generally intended to represent  
turbulent flows. Our results stress that anomalous scaling may 
appear in simple regular (e.g. time-periodic) laminar flows where  
stochasticity appears just as a consequence of the low-dimensional 
deterministic chaos generated by the Lagrangian advection 
dynamics. In addition we have provided numerical evidence for the 
theoretical predictions.  

Chaotic advection is characteristic to most time-dependent flows. 
The linear decay of the advected substance is in fact just the 
simplest prototype of a family of chemical reaction schemes, where 
the local dynamics converges towards a fixed point of the chemical 
rate equations. The local dynamics can also be generated by 
non-chemical processes, e.g. by biological population dynamics in 
the case of plankton advection\cite{PCE}, or by the relaxation of 
the sea-surface temperature towards the local atmospheric 
value\cite{Abraham2}. Inhomogeneities of the chemical sources or 
of other parameters of the local dynamics arise naturally in these 
contexts so that we expect our results to be of relevance in 
biological and geophysical settings. In fact, fractality and 
multifractality have been already observed in these contexts, for 
example in the distribution of stratospheric chemicals (e.g. 
ozone) \cite{Bacmeister,Tuck}, and in sea-surface temperature and  
phytoplankton populations\cite{Seuront}. The structure of these 
fields has been sometimes associated with turbulence of the 
advecting flow. We think that the simple mechanism, able to 
generate complex multifractal distributions, investigated in this 
Paper can be at the origin of some of the structures observed in 
geophysical flows. Further work in this direction could help on 
the interpretation of geophysical data in order to gain 
quantitative information about the processes involved. Laboratory 
experiments seem also to be feasible.  

\acknowledgements 

Helpful discussions with Peter Haynes and Oreste Piro are greatly 
acknowledged. This work was supported by CICYT (Spain, project 
MAR98-0840) and DGICYT (project PB94-1167). Z.N. was supported by 
an European Science Foundation/TAO ({\it Transport Processes in 
the Atmosphere and the Oceans}) Exchange Grant.


\end{document}